# Adaptive Group-based Zero Knowledge Proof-Authentication Protocol (AGZKP-AP) in Vehicular Ad Hoc Networks

*Amar A. Rasheed, Rabi N. Mahapatra, and Felix G. Hamza-Lup. Fellow, IEEE*

*Abstract*—Vehicular Ad Hoc Networks (VANETs) are a particular subclass of mobile ad hoc networks that raise a number of security challenges, notably from the way users authenticate the network. Authentication technologies based on existing security policies and access control rules in such networks assume full trust on Roadside Unit (RSU) and authentication servers. The disclosure of authentication parameters enables user's traceability over the network. VANETs' trusted entities (e.g. RSU) can utilize such information to track a user traveling behavior, violating user privacy and anonymity. In this paper, we proposed a novel, light-weight, Adaptive Group-based Zero Knowledge Proof-Authentication Protocol (AGZKP-AP) for VANETs. The proposed authentication protocol is capable of offering various levels of users' privacy settings based on the type of services available on such networks. Our scheme is based on the Zero-Knowledge-Proof (ZKP) crypto approach with the support of trade-off options. Users have the option to make critical decisions on the level of privacy and the amount of resources usage they prefer such as short system response time versus the number of private information disclosures. Furthermore, AGZKP-AP is incorporated with a distributed privilege control and revoking mechanism that render user's private information to law enforcement in case of a traffic violation.

## I. Introduction

VANET's are special cases of ad hoc networks in which the communicating entities are vehicles, and have variable or no infrastructure. VANETs (see Fig. 1) have emerged for providing comfort and flexible services, cooperative traffic monitoring, alternative routes estimations, real-time assisted navigation, roads closures due to severe weather conditions, collision detection, and avoidance, and access to the global network. This exchange of traffic data among drivers helps enhance passengers' traveling experience over these networks. Many VANETs applications that have been incorporated into the Intelligent Transportation Systems (ITS) [1], [2] [3], [4], [5], [6], [7] networks function on either a peer-to-peer (P2P) communication setup, or via a multi-hop communication setting. The attractive features of VANETs make such networks vulnerable to a wide class of cyber threats that already exist on traditional computer networks [8]. For instance, attacks on the system integrity and availability include message fabrication and delaying, either intentionally or due to hardware malfunctioning. Serious consequences such as injuries and even death may occur due to such types of threats [9], [10], [11], [12], [13], [38]. To alleviate these problems, the development of a functional, reliable, and efficient security framework that integrates critical security features (e.g. authentication, nonrepudiation, and privacy-preserving) is required.

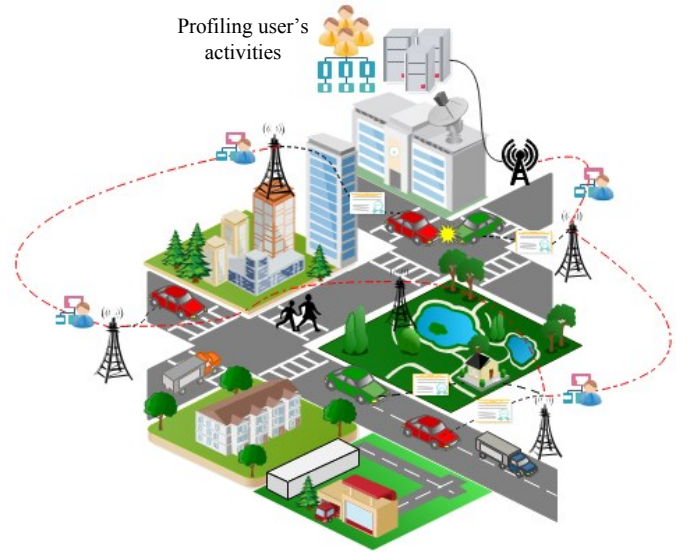

Fig. 1. VANET infrastructure

Although, authentication between onboard units (OBUs) and roadside units (RSUs) plays a crucial role in supporting secure access to VANETs. Several shortcomings have been identified (i) authentication techniques that have been proposed in the past [14], [15], [16], [17] were mainly concerned with providing light-weight yet highly reliable authentication service to vehicles on the road, ignoring user's privacy and tractability (ii) they were developed based on the deployment of security policies that place full trust on the RSUs or the authentication servers in the network. For example, in order to access different services available on VANET, legitimate users have to surrender their authentication parameters to these trusted

Amar Rasheed is an assistant professor with the Computer Science department at Georgia Southern University, Savannah, GA 31419 USA (e-mail: arasheed@ georgiasouthern.edu).

Rabi N. Mahapatra is a professor with the Computer Science and Engineering department at Texas A&M University, College Station, TX USA (email: rabi@cse.tamu.edu).

Felix G. Hamza-Lup is a professor with the Computer Science department at Georgia Southern University, Savannah, GA 31419 USA, (e-mail: fhamzalup@georgiasouthern.edu).

RSUs/Authentication entities. The authentication's parameters may be used by the RSUs or authentication servers to track users' traveling activities on the network. Therefore, concerns about users' privacy may prevent some vehicle owners from joining these systems.

Furthermore, user privacy in these systems is treated as one-model-fits-all. Users operating on these networks are not capable of choosing their own privacy settings. However, privacy is a user-specific concept in the sense that different users may have varying privacy requirements. Moreover, a higher privacy requirement usually results in more computational or communication overhead.

**Our contributions**. Given the conflicting goals of privacy and tractability, and the challenges in designing a light-weight, adaptive privacy-preserving authentication scheme for VANETs, we propose an Adaptive Group-based Zero Knowledge Proof-Authentication Protocol (AGZKP-AP). The protocol is based on the deployment of group-based Zero Knowledge Proof (ZKP) cryptographic system that supports anonymous authentication and distributed revocation between a trusted RSUs and an authorized $OBU_{G_{i,j}}$ group member. Specifically, our main contributions in this paper include:

1. *Authentication protocol.* The proposed protocol provides vehicles' owners with the capability of anonymous authentication over the network. Trusted entities that are part of the VANET system will not be capable of tracking users' activities based on the information they provided during the authentication process.

2. *Distributed privilege control & revoking mechanism.* As authorized OBUs try to access the network, privilege revocation methods will be executed on RSUs to validate if these OBUs are allowed to access the network or not. Misbehaved OBUs that are detected and identified during the network's access time will be broadcasted over the network via a distributed revocation method.

3. *Privacy-preserving threshold defensive scheme.* The proposed scheme is based on ZKP crypto with the support of trade-off options. Users have the option to make critical decisions on the level of privacy and the size of resource usage they prefer such as short system response time versus the amount of private information disclosed. Moreover, the scheme enables users to customize their privacy level settings based on the different services they use on the network.

The paper is organized as follows: Section II presents some of the existing authentication techniques in VANETs and their shortcomings. Section III introduces the background related to the ZKP protocol used during this research effort. Section IV describes an overview of the proposed system's architecture with its supported capabilities. The proposed authentication protocol is presented in section V. Section VI describes the security analysis and the probabilistic model for the proposed protocol. Section VII addresses different threats models against AGZKP-AP, Section VIII studies the behaviors of AGZKP-AP against the ZKP simulator attack. Performance results and the conclusions are presented in sections IX, and X respectively.

## II. EXISTING AUTHENTICATION TECHNIQUES FOR VANETS

A number of authentication protocols have been proposed in the past that support user's anonymity over VANET [32], [33], [34], [35], [36], and [37]. Several protocols are based on cryptographic approaches like verifiable common secret encoding [17]. Using verifiable common secret encoding enables these protocols to provide adaptive anonymity. Verifiable common secret encoding is based on public key cryptography. Privacy-preserving authentication protocols that are based on the above cryptographic approach have some shortcomings when deployed on VANET infrastructures. Heavy cryptographic processing must be performed by each OBUs, leading to high access time.

Raya and Hubaux [4] investigated the privacy issue by proposing a pseudonym based approach using anonymous public keys and the public key infrastructure (PKI), where the public key certificate is needed, giving rise to extra communication and storage overhead. The authors also proposed three credential revocation protocols tailored for VANETs, namely RTPD, $RC^2RL$, and DRP [17], considering that the certificate revocation list (CRL) needs to be distributed across the entire network in a timely manner. All three protocols seem to work well under a conventional PKI. However, the authors also proposed to use frequently updated anonymous public keys to fulfill users' requirement on identity and location privacy. If this privacy preserving technique is used in conjunction with $RC^2RL$ and DRP, the CRL produced by the trusted authority will become very large, rendering the revocation protocols highly inefficient. A lightweight symmetric-key-based security scheme for balancing auditability and privacy in VANETs is proposed in [4]. It bears the drawback that peer vehicles authenticate each other via a base station, which is unsuitable for inter-vehicle communications. Gamage et al. [18] adopted an identity-based (ID-based) ring signature scheme to achieve signer ambiguity and hence fulfill the privacy requirement in VANET applications. The disadvantage of the ring signature scheme in the context of VANET applications, is the unconditional privacy, resulting in the traceability requirement being unattainable. Group signature-based schemes are proposed in [19], [20], [21], where signer privacy is conditional on the group manager. As a result, all these schemes have the problem of identity escrow, as a group manager who possesses the group master key can arbitrarily reveal the identity of any group member. In addition, due to the limitation of group formation in VANETs (e.g., too few cars in the vicinity to establish the group), the group-based schemes [19], [20], [21], [22] may not be applied appropriately. The election of a group leader will sometimes encounter difficulties since a trusted entity cannot be found amongst peer vehicles. Kamat et al. [23], [24] proposed an ID-based security framework for VANETs to provide authentication, nonrepudiation, and pseudonymity. However, their framework is limited by the strong dependence on the infrastructure for short-lived pseudonym generation, which renders the signaling overhead overwhelming. The proposed nonrepudiation scheme enables a single authority to retrieve the identity which may raise the concern of potential abuse. Schemes leveraging pseudonyms in VANETs can also be found in [25], [26] with the revocation feasible in limited settings, and in [27] where the certificate authority maintains a mapping from an identity to the set of vehicle-generated pseudonyms. There are also a number of defense techniques against misbehavior in VANET literature besides those in [4]. An indirect approach via the aid of infrastructure is used in [19] and [23]. The trusted authority (TA) distributes the CRL to the infrastructure points which then take over the TA's responsibility to execute the revocation protocol. The advantage of this approach is that vehicles are not required to download the entire CRL. Unfortunately, the conditional anonymity claimed in [19] and [23] only applies among peer vehicles, under the assumption that the infrastructure points (group manager in [19] and the base station in [23]) are trusted. The infrastructure points can reveal the identity of any vehicle at any time even if the vehicle is honest. The scheme in [28] leverages a single TA to recover the identity of a (possibly honest) vehicle, where revocation issues are not discussed. Recently, Tsang et al. [29] proposed a blacklist-able anonymous credential system for blocking misbehavior without the trusted third party (TTP). The blacklisting technique can be applied to VANETs as: if the vehicle fails to prove that it is not on the blacklist of the current authenticator, the authenticator will ignore the messages or requests sent by this vehicle. Although not proposed specifically for VANETs, the proposal in [29] has a similar claim as ours that the capability of a TTP (network authority in this paper) to recover a user's identity, in any case, is too strong of a punishment and highly undesirable in some scenarios. The downside of this technique is the lack of options to trace misbehaving users since any user in the system (misbehaving or not) will by no means be identified by any entity including the authorities.

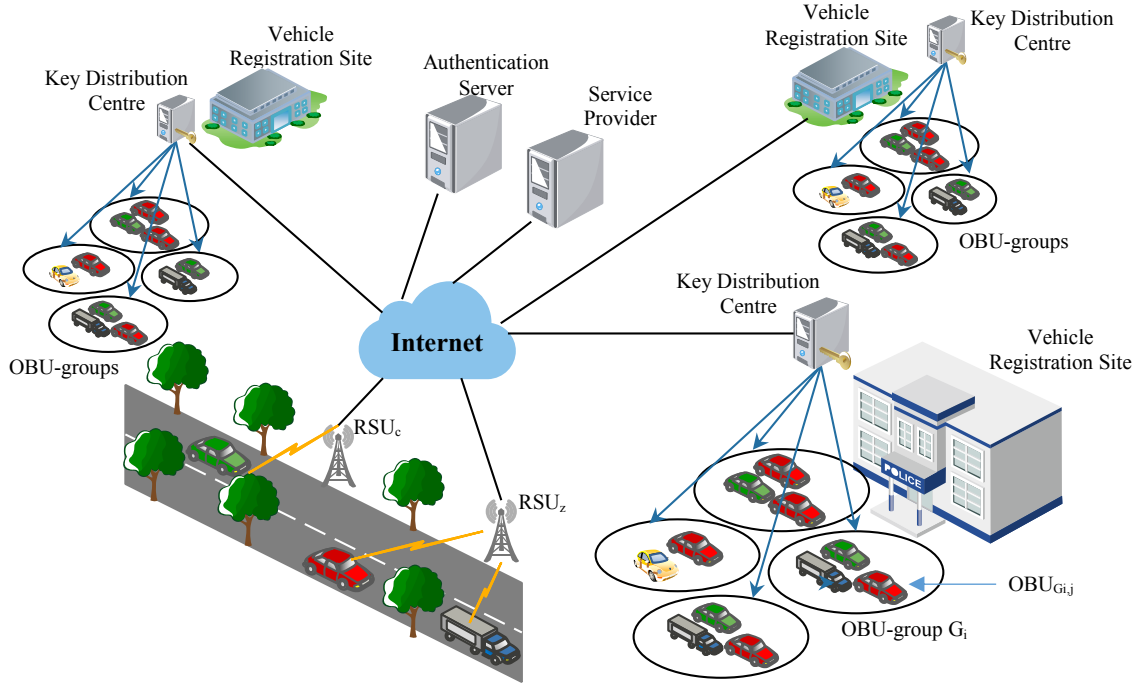

Fig. 2. VANET with the proposed architecture

To the best of our knowledge, authentication protocols that are based on the ZKP approach have not been proposed in the past. The core implementation of our AGZKP-AP was built based on a symmetric cryptosystem and ZKP approach to minimize communication latency and computation time. As opposed to other authentication techniques that adapted the public key infrastructure model, high communication latency and heavy cryptographic processing were observed in such techniques.

### III. PRELIMINARIES

The proposed authentication protocol in this paper is based on the ZKP cryptographic approach. User's anonymity is preserved by incorporating ZKP [31] into the authentication protocol framework. Our authentication scheme is tailored with a distributed revocation method that determines the eligibility of authorized OBUs for accessing the network based on predefined security policies and dynamically updated privileges. The proposed system architecture is depicted in Fig.2.

#### A. Zero Knowledge Proof (ZKP)

ZKP is an interactive identification protocol which enables a prover $P$ to prove his identity polynomially many times to a verifier $V$ without allowing $V$ to misrepresent himself as $P$ to someone else. The proof of identity is either accepted or rejected in real time and as a result, the requested access is granted or ejected. The scheme provides light-weight identification and proves to be suitable for low-end systems with limited processing power [31], like smart card technologies. With carefully preselected parameters, the ZKP scheme is about two orders of magnitude faster than RS-based identification schemes.

The scheme assumes the existence of a trusted center who is involved in publishing a modulus $m$ which is the product of two large prime numbers of the form $4r + 3$. Such moduli are used in a variety of cryptographic applications, and their most useful property is that -1 is quadratic non-residue whose Jacobi symbol is +1 (mod $m$). After publishing $m$, the center can be closed. The ZKP identification scheme relay that a prover $P$ proves to a verifier $V$ that he knows whether a certain number is a quadratic residue or quadratic non-residue (mod $m$) without even revealing a single bit of information.

IV. OVERVIEW OF THE PROPOSED SYSTEM ARCHITECTURE

In this section, we present an overview of the proposed architecture (see Fig. 2) that is tailored to support anonymous authentication over VANETs. The proposed VANET's architecture consists of the following components: (i) roadside units enable vehicles on the road to access different services available on the VANET wirelessly. We consider that RSUs are uniformly distributed over the VANET to allow full coverage over the network (ii) vehicles are equipped with onboard units that enable the wireless exchange of traffic information and users' data between vehicles on the roads and VANET services (iii)Vehicles Registration Sites (VRS) along with onsite Key Distribution Centres (KDC)s are used to generate secret keying information, establish and assign OBU-group members, where secret keys are preloaded on OBUs prior to deployment (iv) Authentication Servers provide OBUs the capabilities of verifying the user credential while accessing different types of services available on VANET (v) Different types of services are available over the network, we consider a network of high-end servers that support different VANET services capabilities for users traveling on the road. In the proposed architecture, several trusted entities including the Key Distribution Centres, the Vehicles Registration Sites take the roles of establishing OBU groups, assigning secret authentication parameters to vehicles and RSU, as well as distributing partially precomputed security primitives over different network entities. As illustrated in Fig. 2, a trusted roadside unit is denoted as $RSU_c$ and an authorized onboard unit is represented as $OBU_{G_{i,j}}$ (a group member), where $G_i$ represents the trusted OBU-group's id and $j$ represents the OBU's id within the trusted group $G_i$. In the proposed protocol, we consider the *zero-trust* model in which users and RSUs do not trust each other. When designing the proposed protocol, we mainly focus on finding solutions to the following list of challenges:

◊ Users' privacy is a big challenge when operating on VANETs' networks, since applying user's private authentication parameters during the OBU-to-RSU authentication, the process might be observed by RSUs or authentication servers that could record and track users' driving activities. User's anonymity is achieved through the employment of the ZKP technique to verify/proof the group-based secrets of the OBUs/RSUs. In order to preserve user' privacy without any server support, we use the group-based anonymous authentication. That is, an OBU only proves its membership within a group of OBUs using group-based secrets, avoiding exposure of its exact identity.

◊ Achieving users' profiling and tractability while keeping anonymity unimpaired is not a trivial task. Networks must support the revocation of misbehaving OBUs and limit their access to the network, and at the same time provide anonymity for others. Providing fast and reliable authentication services for authorized OBUs without compromising users' privacy is critical. We propose a distributed revoking algorithm that is executed autonomously on each RSU deployed on the network. The revoking algorithm dynamically maintains and updates revoking tables with OBUs that need to be excluded from accessing the network.

A. Supported Capabilities

The following includes a list of capabilities that are supported by the proposed protocol.

◊ *Adaptive Anonymity*: In the proposed protocol, users may be concerned with two types of privacy: location and identity privacy (OBU-to-RSU/OBU-to-OBU communications) and the privacy about the service usage patterns (OBU's service requests). The protocol supports multiple anonymity levels, and users are allowed to choose their own privacy setting. Moreover, it enables users to dynamically perform trade-offs between the privacy level and resource utilization according to the users' specific privacy requirements.

◊ *Zero-trust*: Mobile users may want to use different trust policies depending on whether they are communicating with a public or private server (or application). These trust policies include 1) *the full-trust* in which the users trust both types of servers, 2) *the partial-trust* in which the users trust the private or public only, and 3) *the zero-trust* in which the users trust neither of these two types of servers. Previous research work [9], takes the partial-trust policy that trusts some public servers. With these approaches, the authentication requests are sent to some anonymity sever first. Then, the anonymity server sends the anonymized or aggregated requests to other service servers. Therefore, anonymity is achieved at the anonymity server level. In the partial-trust model, the trusted servers have the authentication information, e.g., identity of the mobile user, which can be used to easily track the activities of each individual mobile user based on the spatiotemporal analysis such as the MTT algorithm [12]. In this paper, we focus on the zero-trust model, i.e., the users will trust no RSU or server in the network.

◊ *Traceability* is a required feature and it is supported by the proposed protocol where the identity information can be revealed by law enforcement authorities for liability issues, once accidents or crimes occur.

◊ *Light-weight*: The proposed authentication model was designed to provide quick and reliable secure access to the network by incorporating light-weight cryptographic techniques based on ZKP. Our model is designed to support networks with high speed moving nodes and strict communication range while preserving user privacy. Moreover, the authentication process has a strict real-time response.

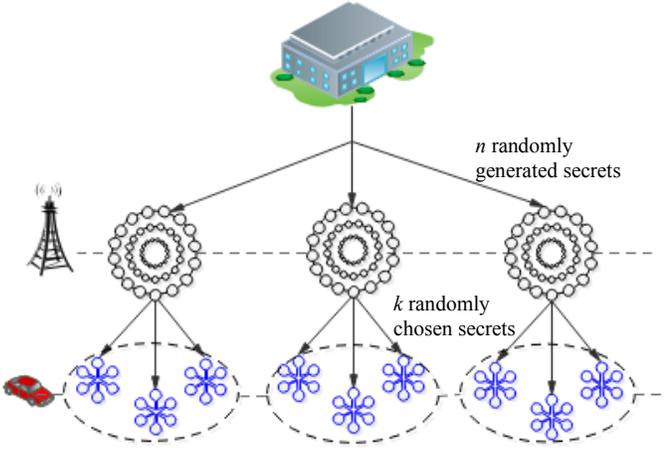

Fig. 3. Key distribution and groups' formation

◊ **Service Differentiation**: Various services will be provided by both private service providers (e.g., automakers and other private companies offering services to the vehicles) and public service providers (e.g., government agencies). Those services need to be differentiated based on the priorities of services and the prices that customers have paid. However, there is a tradeoff between service customization and user anonymity. On one hand, a good resource allocation algorithm should provide customized services for each individual. On the other hand, differentiating services based on specific customer requirements will violate the anonymity requirement of the system. Our model enables users to pick and choose their privacy settings on each service provider that they subscribed for in the network. Based on the user privacy preference he will be able to control his level of exposure on the network.

## V. ADAPTIVE GROUP-BASED ZERO KNOWLEDGE PROOF-AUTHENTICATION PROTOCOL (AGZKP-AP)

Our proposed protocol is composed of:
- ◊ Key Management and OBU-groups Formation
- ◊ Authentication Protocol
- ◊ Distributed Privilege Control Revoking Mechanism

### A. Key Management and OBU-Group Formations

The KDC will employ a Secure Sockets Layer (SSL) technique to digitally sign RSUs public keys. The protocol consists of the following steps:

1. The KDC will generate a pool of certificates and will act as the certification authority for the RSUs.
2. A signed certificate will be distributed to each RSU.
3. RSUs transmit signed certificate as beacon signals to identify their presences over the network.

The signed public keys can be advertised by both the RSU and the KDC to which the RSU belongs at the moment. Authentication process takes place between the RSUs and the OBUs, KDCs were employed only during the keys distribution and the OBU-group formation phase in which it is a one-time process. Digital public certificates were generated by KDCs and preloaded into RSUs at the time of deployment. Our protocol offers mutual authentication capability between an RSU and an OBU and without the involvement of any other VANET network components (e.g. authentication server, KDCs, etc.). Several Key Distribution Centres (KDCs) (see Fig.3) manage the establishment of the OBU-groups and the assignment of group-based secrets for each registered OBU. A KDC generates a set of $q$ OBU-groups with groups ids $\{G_1, G_2, ..., G_i, ..., G_q\}$. For each OBU-group, $G_i$, where $1 \leq i \leq q$. The key generation protocol performs the following tasks:

1. A set $X$ of $n$ secrets $\{S_1, S_2, ..., S_x, ..., S_n\}$ are randomly chosen from the finite field $Z_m$.
2. A subset of $k$ secrets is randomly selected from set $X$ and assigned to each group's member of $G_i$, where $k<<n$.
3. Compute $I_x = \pm S_x^2 (mod\ m)$, where $x \in \{1, 2, ..., n\}$
4. Publish $I_1, I_2, ... I_n$ over the group's members (OBUs), keeping $S_1, S_2, ..., S_n$ private to RSUs.
5. Assigns a unique master secret $S_{G_i}$ which is composed of $k$ secrets $Pr_1, Pr_2, ..., Pr_y, ..., Pr_k$, randomly chosen from the Finite field $Z_m$. The $k$ secrets are preloaded into each OBU-group member.
6. Computes $g_y = \pm Pr_y^2 (mod\ m)$, where $y \in \{1, 2, ..., k\}$
7. Publishes and distributes $g_1, g_2, ..., g_k$ to RSUs, keeping $\{Pr_1, Pr_2, ..., Pr_y, ..., Pr_k\}$ private to OBUs.

Table 1: A list of notations used in AGZKP-AP

| Symbols | Descriptions |
|---------|--------------|
| OBU | Onboard Unit |
| RSU | Roadside Unit |
| KDC | Key Distribution Center |
| $q$ | The total number of randomly generated OBU-groups |
| $G_i$ | The OBU-group id, where $1 \leq i \leq q$ |
| $X$ | Represents the set of secrets in the Finite field $Z_m$, where $m$ is a prime number |
| $n$ | Represents the total number of secrets that are randomly selected and assigned to each RSU |
| $S_x$ | Represents the $x$-th secret in set $X$, where $1 \leq x \leq n$ |
| $k$ | The total number of secrets that are randomly chosen from set $X$ and assigned to an OBU |
| $I_x$ | The whiteness for secret $S_j$ where $1 \leq x \leq n$ |
| $S_{G_i}$ | The $G_i$ OBU-group master key that is composed of $k$ randomly selected secrets |
| $Pr_y$ | The y-th secret of an OBU-group master key where $1 \leq y \leq k$ |
| $g_y$ | The whiteness for secret $Pr_y$ where $1 \leq y \leq k$ |
| $OBU_{G_i,j}$ | An OBU member with id, $j$ within group $G_i$ holds a copy of the master secret $S_{G_i}$ |
| $\mu$. | The total number of ZKP terms generated during each authentication occurring between an OBU and RSU |
| $\alpha$ | A privacy performance metric that limits the number of ZKP terms needed to be verified for successful authentication. |
| $Cert(Pub_y)$ | RSUs public certificates |

In the proposed authentication protocol, *m* is public and preloaded into the RSUs and OBUs during the vehicles' registration process. Secrets keys $\{S_1, S_2, \ldots, S_x, \ldots, S_n\}$, and $\{Pr_1, Pr_2, \ldots, Pr_y, \ldots, Pr_k\}$ are kept private to RSUs and OBUs respectively. The $S_x$ or $Pr_y$ (which are witnesses to the quadratic residuosity character of the $I_x$ or $g_y$ respectively) are effectively hidden by the difficulty of extracting square roots *mod m*, and thus prover **P** can establish his identity by proving that he knows these $S_x$ or $Pr_y$. By allowing $I_x$ and $Pr_y$ to be either plus or minus a square modulo a Blum integer, the protocol make sure that $I_x$ and $Pr_y$ can range over all the numbers with Jacobi symbol +1 *mod m* and thus $S_x$ and $Pr_y$ exist (from the verifier *V* point of view) regardless of $I_x$ and $Pr_y$ character as required in the ZKP. OBU-group members can be identified based on their assigned private secrets set which are kept private at the KDCs, No other entities of the VANET like RSU and authentication servers have access to such information. We assume that authentication servers, RSUs and KDCs are not cooperating during the OBU-to-RSU authentication process. Private secrets are preloaded into each vehicle's OBU during registration. The proposed authentication model was inspired from the *verifiable common secrets encoding* algorithm. Common verifiable secrets integrated with a ZKP cryptography solution are used to solve the problem RSU-to-OBU anonymous authentication, preventing an OBU from releasing its private authentication information to RSU/Authentication servers during this process.

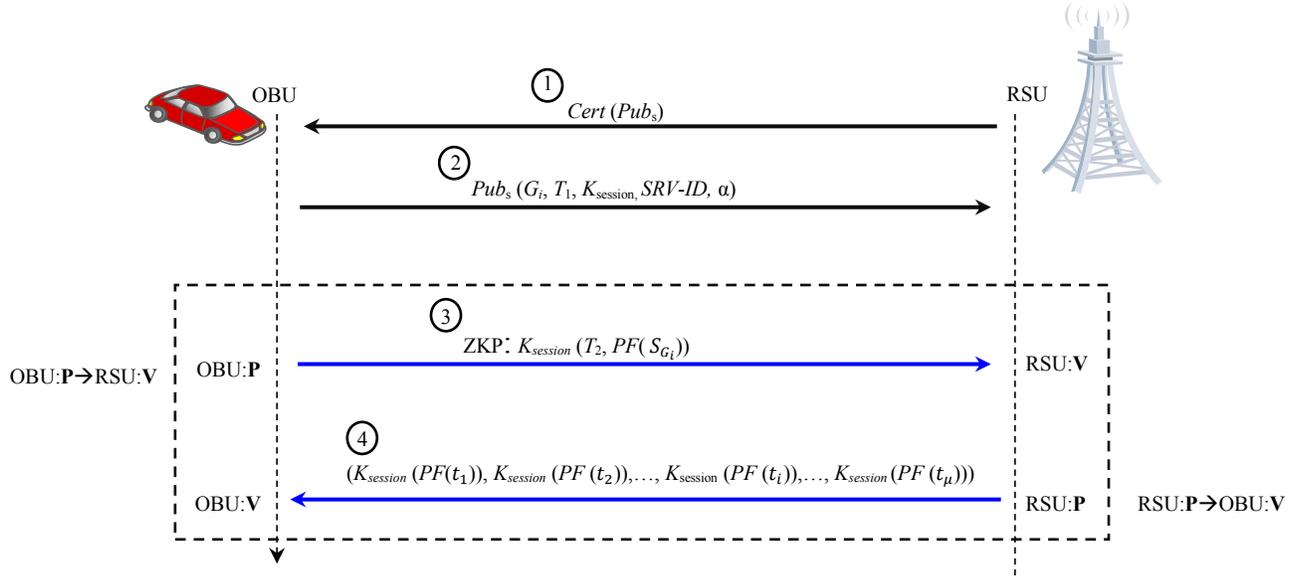

Fig. 4. The proposed authentication protocol (AGZKP-AP)

*B. Authentication Protocol*

In the proposed protocol, anonymity is achieved via the implementation of a two-way ZKP cryptographic protocol. We have adopted a two-way verification and proving crypto approach based on ZKP, where RSU acts as the verifier and OBU as the prover during the OBU-to-RSU authentication phase. RSUs' and OBUs' proofs were encrypted/decrypted using symmetric encryption technique, AES128. We employed a hybrid cryptosystem that uses the public key infrastructure to securely distribute session keys between the RSU and the OBU.

The proposed protocol relied on AES crypto to facilitate the quick exchange of secure authentication parameters between RSUs and OBUs. Various privacy's models are supported by the proposed protocol. We applied a user-centric approach when defining the amount of private information disclosure during the authentication process. In the proposed protocol, we consider the use of a privacy performance metric α that is incorporated into the design of AGZKP-AP. We use this performance metric to limit the number of ZKPs needed to be verified for successful authentication. During the authentication process, AGZKP-AP provides users with the capabilities of the dynamic allocation of resources in terms of communication overheads, latency, and the number of information leakages. α is provided to control the number of private authentication parameters that are exposed on the network. The protocol supports multiple anonymity levels {α=1, α=2, α=3, α=4, α=5}, and users are allowed to choose their own privacy setting. As α increases, the amount of the private information being disclosed over the network increases, highest level of users' anonymity is achieved setting α=1. Moreover, AGZKP-AP enables users to dynamically perform trade-offs between privacy level and resources utilization according to the users' specific privacy requirements. Based on user preferences and the type of service being requested, a privacy parameter α is proposed by the requester. Both the service provider and the requester must mutually agree on the privacy value before processing any authentication requests. For example, $OBU_{G_i,j}$ (**P**rover) submits proof of membership to an RSU (**V**erifier), that is a member of group $G_i$ and it holds a copy of the master secret $S_{G_i}$. That is, an OBU only proves its membership within a group of OBUs using a group-based secret, avoiding its identity exposure to the RSUs. During the RSU-to-OBU authentication phase, the **V**erifier (OBU) will engage in *μ* ZKP sessions with the **P**rover (RSU), where *μ* is a performance metric incorporated into the protocol design, such metric can be used to trade off reliability for latency. It will define a threshold value for the maximum number of ZKP proofs that need to be verified. RSU submits $\mu$ proofs of knowledge indicating that it holds a copy of the OBU-group-based secrets $\{S_1, S_2,…, S_n\}$. An OBU controls the amount of private information being leaked during the authentication process by limiting the number of required ZKP verifications to α, where α≤$\mu$. In order for an RSU to be authenticated by an OBU, the OBU must be able to successfully verify at least *α* RSU-proofs. To avoid the exposure of the user's identity during authentication, verification of the OBU's private secrets are performed at the OBU level and not at the RSU level. An OBU will be able to anonymously authenticate an RSU and without leaking extra information that could lead to user identity theft. A high-level system model with the AGZKP-AP algorithm is depicted in Fig. 4 and Fig. 5:

1. RSU→OBU: Cert (Pubs). RSU announces its presence periodically with its digital public certificate. OBU→RSU: $Pub_s$ ($G_i$, $T_1$, $K_{session}$, SERV-ID, α). OBU constructs a message with its group identifier $G_i$, current time $T_1$, a session key $K_{session}$, the requested service's id SERV-ID, and a user-selected privacy parameter α. It then encrypts the message with the RSU's public key $Pub_s$.

2. Requests submitted by authorized users will be verified with the services providers to determine if a given request with a privacy parameter α is allowed through the network or not. In order for a user to access a service on the network, both the requester and the service provider must establish a mutual agreement on the level of privacy used.

3. OBU:**P**→RSU:**V**. RSU and OBU initiate the ZKP protocol, OBU acts as a prover and sends a proof of knowledge $PF(S_{G_i})$ generated at time $T_2$. The proof is encrypted using the session key $K_{session}$ ($T_2$, $PF(S_{G_i})$). RSU will verify the OBU's proof to achieve OBU-to-RSU authentication.

4. RSU:**P**→OBU:**V**. OBU and RSU engage in $\mu$ ZKP sessions. RSU submits $\mu$ encrypted proofs of knowledge ($K_{session}$ ($PF(t_1)$), $K_{session}$ ($PF(t_2)$),…, $K_{session}$ ($PF(t_i)$),…, $K_{session}$ ($PF(t_\mu)$)), where each $PF(t_i)$ is computed by randomly choosing $k$ secrets from the OBU-group-based secrets ($S_1, S_2,…, S_n$) in $G_i$. This dynamic generation of the $\mu$ proofs is determined by the service requester ($OBU_{G_{i,j}}$), such that no two proofs in the above setting consists of the same $k$ secrets sequence.

5. OBU decrypts these $\mu$ RSU's proofs and confirms anonymity. Upon successful decryption and verification, it constructs a reply message with the value α. In the proposed protocol, the value of the privacy parameter α is used to determine the minimum numbers of RSU's proofs that an OBU must be able to verify during the OBU-to-RSU authentication phase. We use α as a performance metric that allows users to trade-off between authentication reliability and speed.

## C. Protocol's Security Model

User's anonymity is achieved in AGZKP-AP by employing the following ZKP-based technique:

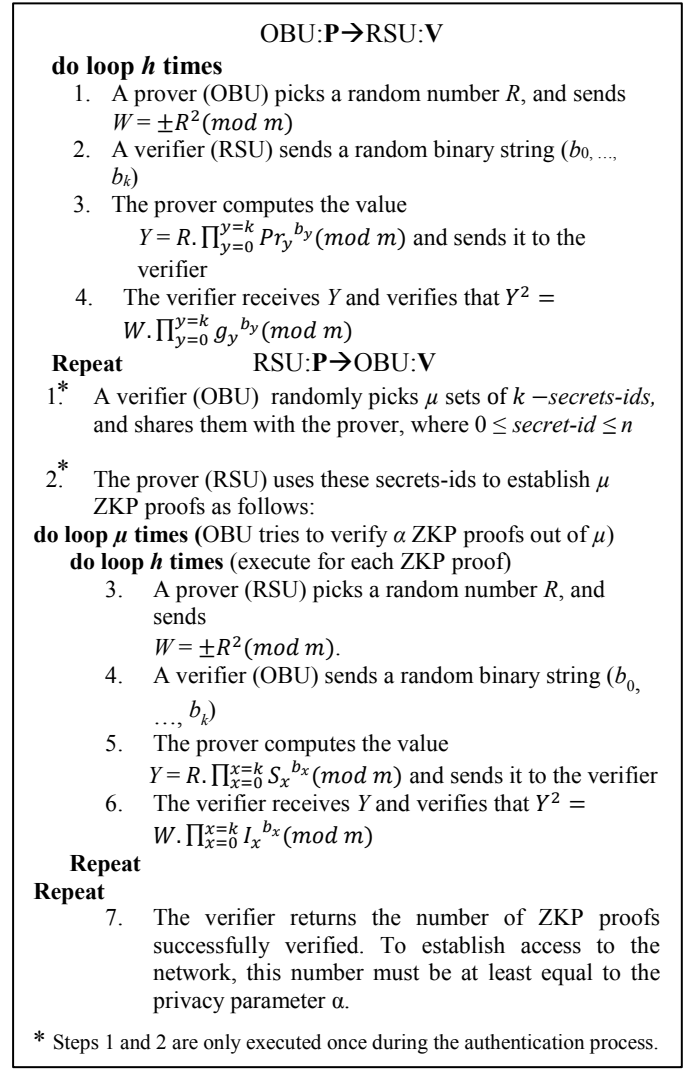

OBU:**P**→RSU:**V**
**do loop *h* times**
1. A prover (OBU) picks a random number *R*, and sends $W = \pm R^2 (mod\ m)$
2. A verifier (RSU) sends a random binary string ($b_0, …, b_k$)
3. The prover computes the value $Y = R \cdot \prod_{y=0}^{y=k} Pr_y^{b_y} (mod\ m)$ and sends it to the verifier
4. The verifier receives *Y* and verifies that $Y^2 = W \cdot \prod_{y=0}^{y=k} g_y^{b_y} (mod\ m)$

**Repeat**    RSU:**P**→OBU:**V**
1.* A verifier (OBU) randomly picks $\mu$ sets of $k$ −*secrets-ids*, and shares them with the prover, where $0 \le secret\text{-}id \le n$

2.* The prover (RSU) uses these secrets-ids to establish $\mu$ ZKP proofs as follows:
**do loop *$\mu$* times** (OBU tries to verify α ZKP proofs out of $\mu$)
  **do loop *h* times** (execute for each ZKP proof)
3. A prover (RSU) picks a random number *R*, and sends $W = \pm R^2 (mod\ m)$.
4. A verifier (OBU) sends a random binary string ($b_0, …, b_k$)
5. The prover computes the value $Y = R \cdot \prod_{x=0}^{x=k} S_x^{b_x} (mod\ m)$ and sends it to the verifier
6. The verifier receives *Y* and verifies that $Y^2 = W \cdot \prod_{x=0}^{x=k} I_x^{b_x} (mod\ m)$

**Repeat**
**Repeat**
7. The verifier returns the number of ZKP proofs successfully verified. To establish access to the network, this number must be at least equal to the privacy parameter α.

* Steps 1 and 2 are only executed once during the authentication process.

Fig. 5. AGZKP-AP algorithm

### C.1 Scenario: Two OBUs accessing VANET via RSU$_j$

To illustrate the visibility of the proposed protocol in terms of preserving the privacy of two OBUs trying to authenticate to the VANET via the same RSU. We consider the case of two OBUs, $OBU_{G_{1,a}}$ and $OBU_{G_{2,b}}$ from groups $G_1$ and $G_2$ respectively, accessing the network via RSU$_j$. Group $G_1$ is assigned the set of secrets X: $\{S_{11}, S_{12},…, S_{1n}\}$, Group $G_2$ is assigned the set of secrets Y: $\{S_{21}, S_{22},…, S_{2n}\}$. As described early, RSUs are preloaded with the randomly generated sets of OBU-groups based secrets, including the sets of secrets for groups $G_1$ and $G_2$. Two sets of witnesses $W_{G_1}$: $\{I_{11}, I_{12},…, I_{1n}\}$, and $W_{G_2}$: $\{I_{21}, I_{22},…, I_{2n}\}$ are securely computed by the KDC and stored in the OBUs, $OBU_{G_{1,a}}$ and $OBU_{G_{2,b}}$ respectively. In addition to the witnesses sets, unique group-based master keys $S_{G_1}$ and $S_{G_2}$ are preloaded into $OBU_{G_{1,a}}$ and $OBU_{G_{2,b}}$ respectively. Meanwhile, witnesses sets for the group-based keys, $S_{G_1}$ and $S_{G_2}$ are stored in every RSU connected to the network including RSU$_j$. The following list of interactions take place between the two OBUs and RSU$_j$:

1. The discovery of RSU$_j$ by OBUs using the PKI approach
   - ❖ $OBU_{G_1,a}$, transmits a message encrypted with RSU$_j$ public key. The encrypted message comprised of a timestamp, a randomly generated session key K1, the group's id, $G_1$, the requested service's id SERV-ID, and a user-selected privacy parameter α (e.g. α =2).
   - ❖ $OBU_{G_2,a}$, transmits a message encrypted with RSU$_j$ public key. The encrypted message comprised of a time stamp, a randomly generated session key K2, the group's id, $G_2$, the requested service's id SERV-ID, and a user-selected privacy parameter α (e.g. α =2).
   - ❖ RSU$_j$ tags each received session keys, K1 and K2 with a unique id, ID$_{K1}$ and ID$_{K2}$ respectively. IDs are randomly generated for each authentication session. We used these dynamically generated IDs to identify OBUs' session keys when applying encryption and decryption.

2. $OBU_{G_1,a}$ and $OBU_{G_2,b}$ in this step act as proofers and RSU$_j$ acts as a verifier
   - ❖ $OBU_{G_1,a}$ constructs a proof of knowledge K1(Timestamp, $PF(S_{G_1})$) encrypted with key, K1 and sends it to RSU$_j$.
   - ❖ $OBU_{G_2,b}$ constructs a proof of knowledge K2(Time stamp, $PF(S_{G_2})$) encrypted with key, K2 and sends it to RSU$_j$.
   - ❖ RSU$_j$ make uses of the witnesses sets, $W_{G_1}$, $W_{G_2}$ it possess to verify that both OBUs hold the right sets of secrets. $OBU_{G_1,a}$ and $OBU_{G_2,b}$ proofs are individually verified by RSU$_j$.

3. RSU$_j$ acts as a proofer, and $OBU_{G_1,a}$, $OBU_{G_2,b}$ act as verifiers
   - ❖ RSU$_j$ constructs $\mu$ encrypted proofs of knowledge (K1 $(PF(t_1))$, K1 $(PF(t_2))$,…, K1 $(PF(t_\mu))$). RSU$_j$ sends the encrypted proofs to $OBU_{G_1,a}$. Proofs are computed by randomly choosing $k$ secrets from the OBU-group-based secrets ($S_{11}, S_{12},…, S_{1n}$) in $G_1$. This dynamic generation of the $\mu$ proofs are determined by the service requester ($OBU_{G_1,a}$) and exchanged securely with RSU$_j$, such that no two proofs in the above setting consists of the same $k$ secrets sequence.
   - ❖ RSU$_j$ constructs $\mu$ encrypted proofs of knowledge (K2 $(PF(t_1))$, K2 $(PF(t_2))$,…, K2 $(PF(t_\mu))$). RSU$_j$ sends the encrypted proofs to $OBU_{G_2,b}$. Proofs are computed by randomly choosing $k$ secrets from the OBU-group-based secrets ($S_{21}, S_{22},…, S_{2n}$) in $G_2$. This dynamic generation of the $\mu$ proofs are determined by the service requester ($OBU_{G_2,b}$) and exchanged securely with RSU$_j$, such that no two proofs in the above setting consists of the same $k$ secrets sequence.
   - ❖ $OBU_{G_1,a}$ and $OBU_{G_2,b}$ decrypt the RSU$_j$ proofs. They achieved anonymous authentication by validating α-proofs (α ≤ $\mu$), the two OBUs use their witness data sets, $W_{G_1}$ and $W_{G_2}$ to verify the RSU$_j$ identity, without the release of private keying information.

Identification of $OBU_{G_1,a}$ and $OBU_{G_2,b}$ take place on the group-level to preserve OBUs privacy.

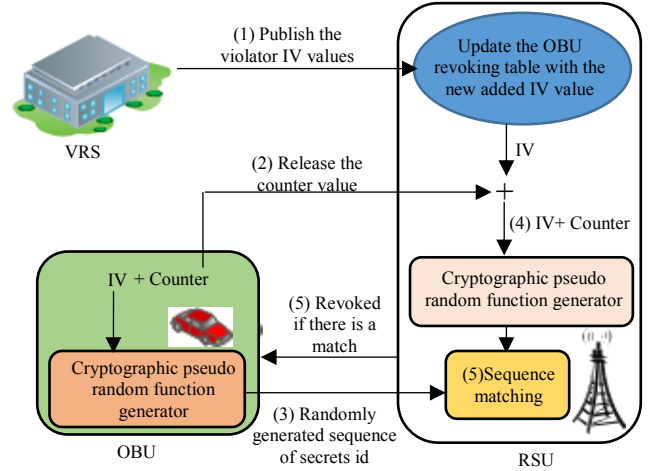

Fig. 6. Privilege control and revoking mechanism

*D. Distributed Privilege Control and Revoking Mechanism*

At the time of OBUs deployment, each OBU is preloaded with a unique 64-bit initialization vector (IVs) assigned by the vehicle manufacturer. OBUs surrender their initialization vectors values to the VRS during the key management and OBU-group formation step. To protect user's privacy, no trusted entities on the network have access to IVs values except VRS and KDCs. We employed counter mode encryption to generate an unpredictable sequence of secrets id each time an OBU accessing the network. Prior to joining the network, each OBU initialized its counter value to zero. As an OBU tries to join the network, it increments the counter value by one after each successful authentication, this value is securely shared with RSUs over the network in the case of identifying and isolating a violator. In this protocol (see Fig. 6), users relinquish their privacy when they attempt an act that violates the network access rules and policies. OBUs identification credentials will be rendered and made available for trusted entities on the network.

Tractability and user profiling is maintained by the network through the utilization of a distributed privilege control and revocation scheme. A violator of the VANET network policy will be identified and revoked from the network at the time of accessing the network. For example, in the case of a violator trying to access a particular service available on VANET, the proposed scheme populates the network with the violator IV value. All trusted entities including RSUs will obtain a copy of the IV and update their OBU revoking tables. OBU revoking tables are maintained by RSUs and are dynamically updated by VRS. When an $OBU_{G_i,j}$ violator is identified by the network, an RSU will modify its assigned unique master secret $S_{G_i}$ to some garble values preventing it from future access via its $G_i$ OBU-group membership. Our distributed privilege control and revoking mechanism employ a cryptographic pseudo-random function generator [30] that uses (IV + counter) as a seed to generate unpredictable sequences for each authorized OBU. OBUs' sequences are composed of $(k \times \mu)$ secrets' ids randomly chosen from a pool of *n* available ids. We use seed values to track and identify network violators. A pattern matching algorithm was used to reconstruct the OBU sequence on the fly in which it's compared to the received OBU sequence. In the case of a match, the RSU will attempt to isolate the OBU from the network.

## VI. SECURITY ANALYSIS FOR AGZKP-AP

This section provides probabilistic models for the proposed protocol.

### A. Probabilistic Modeling of the OBU-to-RSU Authentication scheme.

As previously discussed, during the OBU-to-RSU authentication process, an OBU (prover) and an RSU (verifier) execute the ZKP protocol once, where an OBU tries to prove his OBU-group membership to the RSU. We have estimated the probability of an OBU cheater where an OBU can easily cheat a verifier (RSU) with probability $P_c$:

$$P_c = \left(\frac{1}{2^{kh}}\right)$$

A cheater needs to guess a random binary vector ($b_0, b_1, \ldots, b_k$) with a probability of $2^{-k}$ per iteration, preparing $W = \pm R^2 / \prod_{i=0}^{i=k} g_i^{b_i} (mod\ m)$ in step 1, & providing $Y = R$ in step 3.

### B. Probabilistic Modeling of the RSU-to-OBU Authentication Scheme

The main difference between the RSU-to-OBU and OBU-to-RSU authentication schemes is the required number of ZKP proofs that need to be established during the verification process. We defined a threshold value $\mu$ that determines the minimum number of ZKP proofs a verifier needs to generate based on a random subset of ($k \times \mu$) secrets chosen by the verifier (OBU) from the pool of *n* available secrets. We assume that all $\mu$ proofs are uniquely established by the verifier such that no two proofs, $PF(t_i)$ and $PF(t_j)$ are equal. The estimated probability of an RSU cheater is:

$$P_\mu = \left(\frac{1}{2^{kh} \cdot \binom{n}{k}}\right)^\mu$$

### C. User's Privacy and OBU Identification Information Leakage

During the RSU-to-OBU authentication, RSUs and OBUs engage in a selection process, where an OBU randomly establishes $\mu$ sets of *k-secret-ids,* and shares them with the prover. We have estimated the probability $P_L$ of having a generated set that leaks the OBU identification information as:

$$P_L = \frac{\mu}{\binom{n}{k}}$$

The proposed protocol offers a tradeoff between user's anonymity and the strength of verifying an RSU by OBU.

### D. Probabilities Estimation for False Authentication

False authentication occurs in the proposed protocol when two OBUs from the same group establish similar $\mu$ ZKP poofs with an RSU. As illustrated in section C, similarities occur when two OBUs in the same group submit identical $\mu$ sets, each of *k-secret-ids* to the RSU. The latter leads to the establishment of similar $\mu$ ZKP proofs. We computed the following probabilities:

- The probability $q$ that two OBUs used the same sequence of $\mu$ ZKP proofs for authentication and RSU verification.

$$q = \left(\frac{1}{2^{kh} \cdot \binom{n}{k}}\right)^\mu$$

- The probability $q$ that *x* OBUs used the same sequence of $\mu$ ZKP proofs for authentication and RSU verification.

$$q_x = \left(\frac{1}{2^{x(k-1)} \cdot \binom{n}{k}^{x-1}}\right)^\mu$$

### E. Probability Estimation for Missed OBU Revocation

An OBU violator is missed with a probability $p$ by our protocol's revocation scheme when there are two OBUs with the same OBUs' sequences. As presented in section III, OBUs' sequences are composed of ($k \times \mu$) secret ids randomly chosen from a pool of *n* available ids. The probability to have two OBUs with the same sequence is computed as:

$$p = \frac{1}{\binom{n}{k} \times \left(\binom{n}{k} - 1\right) \times \left(\binom{n}{k} - 2\right) \times \ldots \ldots \times \left(\binom{n}{k} - \mu\right)}$$

## VII. THREAT MODELING FOR AGZKP-AP

In this section, we describe a threat model (Fig. 7) that enables a simulator to circumvent all security measures implemented in the proposed protocol. The attack is orchestrated by (i) employing a passive observer that records steps 3, 4, and 6 of the communication during each stage of the RSU-to-OBU authentication scheme (ii) For each set of $k$ randomly chosen secrets, spoofed $Y$ and $W$ values along with their corresponding binary strings are recorded and used to construct a single ZKP simulator. The ZKP simulators are used by the attacker to spoof an RSU. Each constructed ZKP simulator is capable of establishing a single ZKP proof. In the proposed protocol, $\mu$ ZKP proofs are constructed based on a random selection process of $\mu$ sets, each with $k$-secrets. Based on this random selection process, $\binom{n}{k}$ ZKP proofs can be developed. Therefore for an attack to succeed with a probability close to 1, $\binom{n}{k}$ ZKP simulators are required with a memory requirement of $2^{2k+6} \times \binom{n}{k}$ bytes (assuming each $Y$ is a 64-bits value). In this work, our simulator can be represented as a matrix of size $(2^k \times 2^k)$, where $k$ is the length of the binary string (see Fig. 8 and Fig. 9).

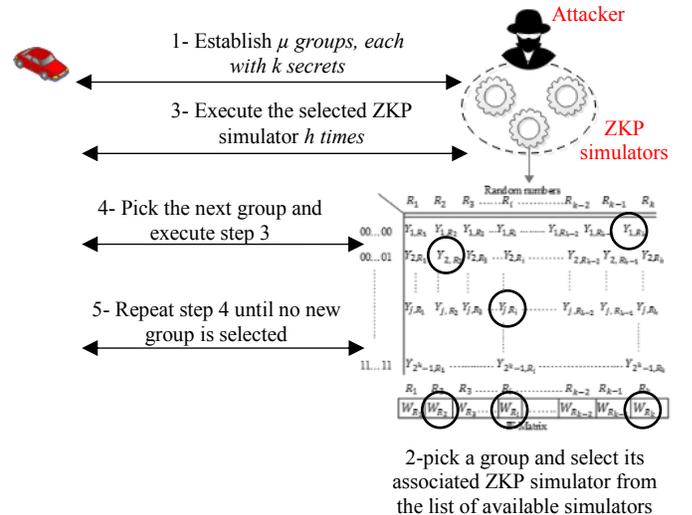

Fig. 7. Threat models

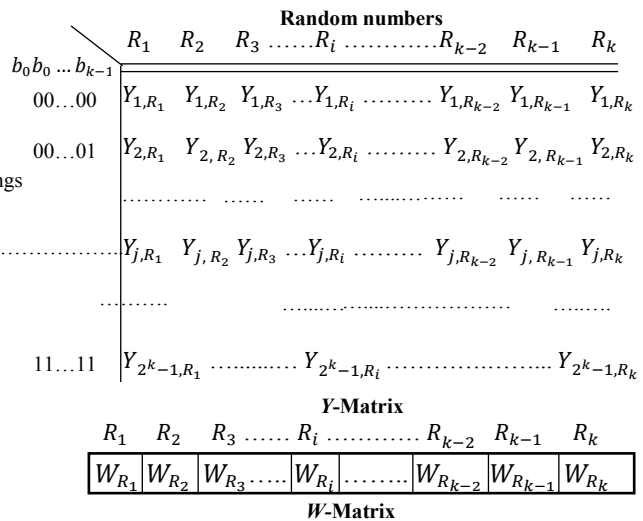

Fig. 8. A single ZKP simulator
Fig. 9. Implementation of the threat model

## VIII. COUNTERMEASURE AGAINST ZKP SIMULATOR ATTACKS

In this section, we address one weakness in the proposed AGZKP-AP and introduce a defensive mechanism against ZKP simulator attacks. ZKP protocols are susceptible to replay attacks. A ZKP simulator (cheater) can be constructed by observing and recording all ZKP's communications. Since ZKP proofs are generated based on choosing a random binary string of length $k$, and random number $R$, a ZKP simulator attack is inexpensive to launch. It requires only $2^{2k+6} \times \binom{n}{k}$ in terms of memory cost. To close this security gap, we proposed a novel technique that modifies internally how ZKP proofs are computed. We introduce a new ZKP method, where proofs are

---

**Attacker: P→OBU:V**

1. A verifier (OBU) randomly picks $\mu$ sets of $k$ −*secrets-ids*, and shares them with the prover, where $0 \le$ *secret-id* $\le n$
2. The prover (Attacker) uses these secrets-ids to identify the correct $\mu$ ZKP simulators for execution.

**do loop $\mu$ times (For each spoofed ZKP proof, pick a simulator )**
 **do loop $h$ times (execute a simulation for each spoofed ZKP proof)**
 3. The attacker picks a value $W_{R_i}$ from the $W$-matrix, and sends it to the verifier
 4. A verifier (OBU) sends a random binary string $S$ ($s_0, ..., s_k$)
 5. The attacker picks a value $Y_{S,R_i}$ from the $Y$-Matrix and sends it to the verifier
 6. The verifier receives $Y_{S,R_i}$ and verifies that $(Y_{S,R_i})^2 = W_{R_i} \cdot \prod_{i=0}^{i=k} I_i^{s_i} (mod\ m)$
**Repeat**
**Repeat**
 7. The verifier returns the number of ZKP proofs successfully verified. To establish access to the network, this number must be at least equal to the privacy parameter α.

\* Steps 1 and 2 are only executed once during the authentication.

---

computed based on evaluating a shared polynomial $F(x)$ of degree $k$. To compute a ZKP proof that is composed of $k$ secrets, (i) a shared polynomial of degree $k$ is constructed between a proofer and a verifier, where $k$ polynomial coefficients are securely computed over the finite field $Z_Q$ at both ends using a cryptographic hash function. The shared polynomial $F(x)$ is represented as:

$$F(x) = \sum_{k=0}^{k-1} a_k\, x^{k \cdot b_k}$$

The modified version of the ZKP protocol implements the following list of interactions.

1. A prover (RSU) picks a random number $R$, and sends $W = \pm R^2 (mod\ m)$.
2. The verifier (OBU) and the proofer securely construct the shared polynomial $F(x)$ independently using a secure cryptographic hash function.
3. The verifier (OBU) sends a random binary string ($b_0, ..., b_k$)
4. The prover computes the following values:

$$g(x) = \prod_{i=1}^{k} \sum_{k=0}^{k-1} a_k\, S_i^{2k \cdot b_k}$$

 $Y = (R^2 \cdot g(x)) mod\ m$ and sends it to the verifier
5. The verifier receives $Y$ and verifies that $Y \cdot Y' = W$, where $Y' = (1/ \prod_{i=1}^{k} \sum_{k=0}^{k-1} a_k\, I_i^{k \cdot b_k})\ mod\ m$
6. Steps 1 through 5 are repeated $h$ times.

## IX. PERFORMANCE RESULTS OF AGZKP-AP

We evaluate the resiliency of the proposed protocol against an RSU cheater. As illustrated in section VI, an RSU cheater needs to guess a random binary string with a probability $2^{-k}$ per iteration, prepares $W = \pm R^2 / \prod_{i=0}^{i=k} g_i^{b_i} (mod\ m)$ in step 1, and computes $Y = R$ in step 3. Figures 10a & 10b illustrate how AGZKP-AP behaves with different tradeoff options. In Fig. 10a, we compute the probability $P_\mu$ with different $h$ iterations per ZKP proof, where the value $h \in \{4,5,6,8\}$.

In our protocol, $h$, and the required number of ZKP proofs to verify an RSU, $\mu$ are utilized to estimate the resiliency of AGZKP-AP. At the same time these performance metrics offer a tradeoff between user's anonymity and computation costs.

Employing a higher number of $h$ iterations per ZKP improves the protocol resiliency against a cheater, but at the same time introduces high communication overhead between an OBU and an RSU. Time-sensitive VANET services, for example, Emergency Response System (ERS) requires low latency and high data reliability to save lives. AGZKP-AP offers such capability by dynamically modifying the internal structure of ZKP to accommodate different types of service demands.

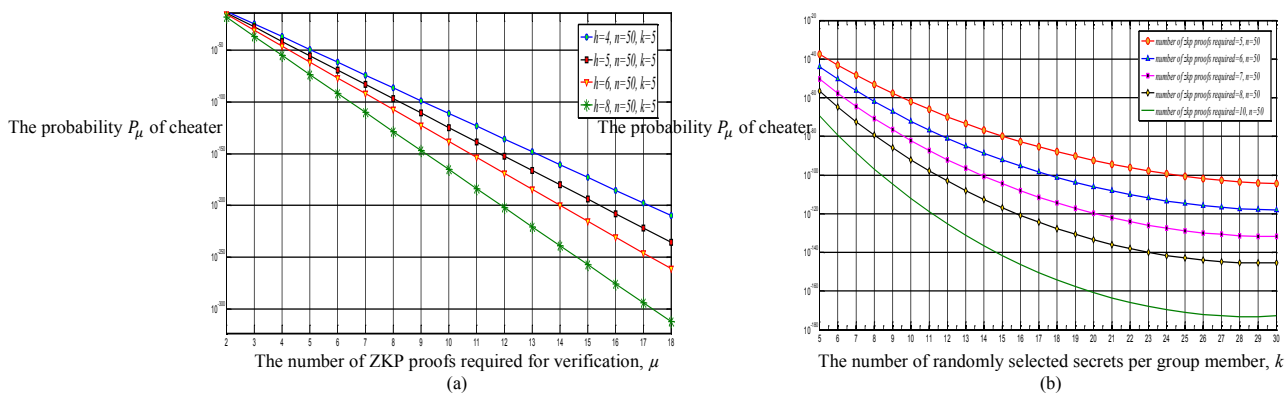

Fig. 10. The probability $P_\mu$ of an RSU cheater

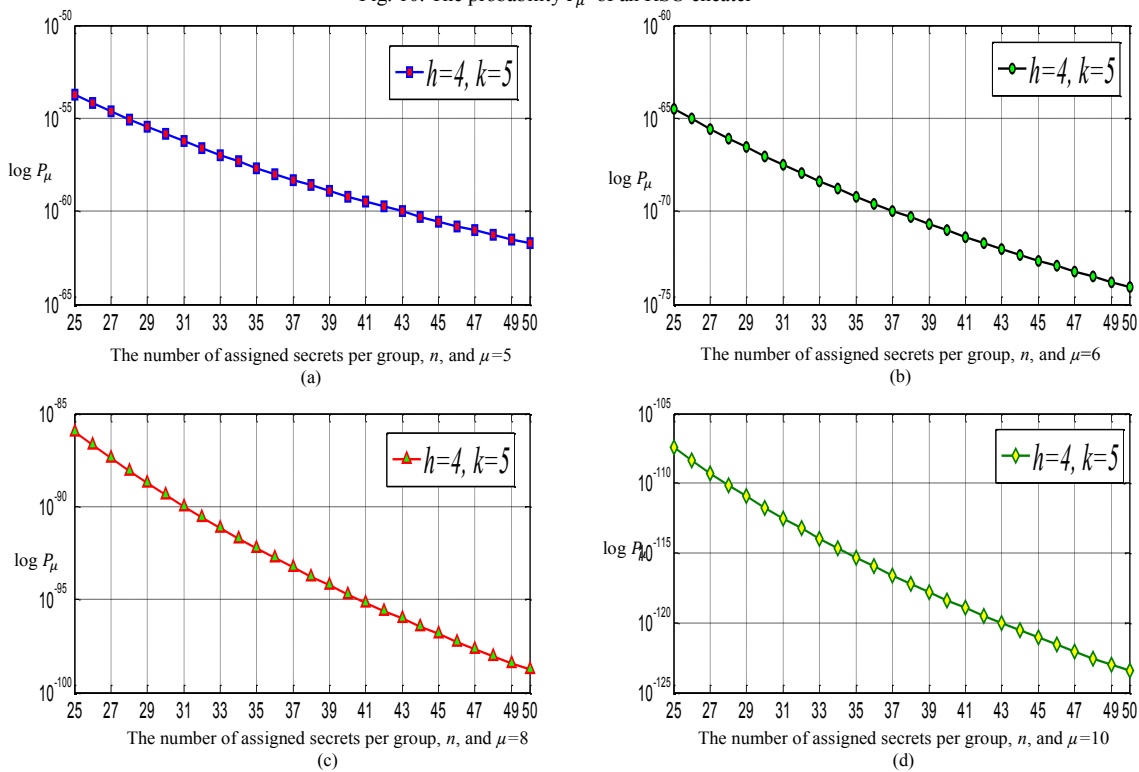

Fig. 11. The probability $P_\mu$ of an RSU cheater versus the number of assigned secrets per

Fig. 10b, presents the resiliency of the proposed protocol with various μ ZKP proofs used for verification, as the value of μ increases from 5 ZKP proofs to 10 ZKP proofs, the probability of cheater decreases since it takes more effort for an attacker to correctly guess all the randomly picked binary strings for each of the μ ZKP poofs required during the verification process. The attacker needs to correctly solve each of the μ ZKP poofs required for verification. Also, a higher number of secrets, k being chosen per group member has a direct impact on the probability of having a successful cheater. In Fig 11, we further analyze the tolerance of AGZKP-AP against an RSU cheater, by estimating the probability of having a successful cheater with a fixed number of assigned k secrets per group member and a fixed number of h iterations per ZKP (k =5, h=4). Two critical performance metrics, $\mu$ and $n$ are used to capture the behavior of the AGZKP-AP during an RSU cheater attack. As the number of assigned secrets, $n$ increases, the probability $P_\mu$ decreases exponentially. Fig. 11 shows the probability of having a successful RSU cheater with various values for $\mu$.

User's privacy is controlled by the amount of private information being used during the RSU-to-OBU authentication scheme. Information leakage is estimated by evaluating the probability, $P_L$ of having at least one ZKP proofs that carry OBU identification information ($k$ randomly selected secrets per group member). Fig. 12 shows the probability, $P_L$ with

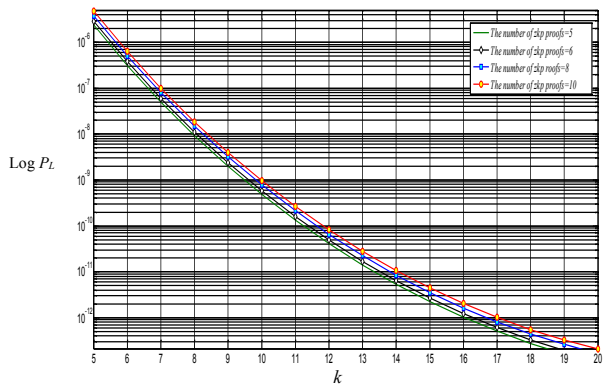

Fig. 12. The probability $P_L$, $n = 50$

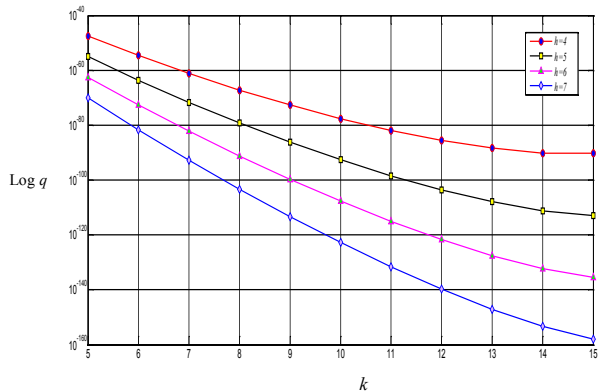

Fig. 13. The probability of false authentication, $n = 15$, $\mu = 5$

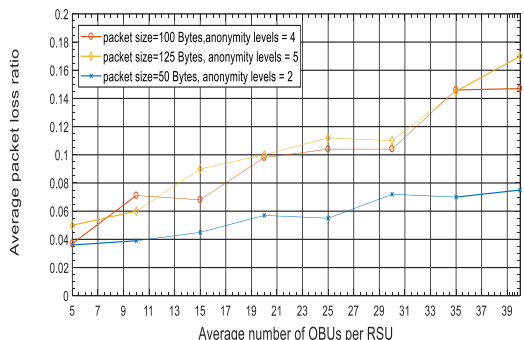

(a)

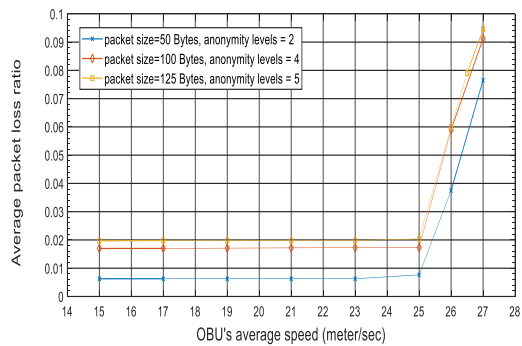

(b)

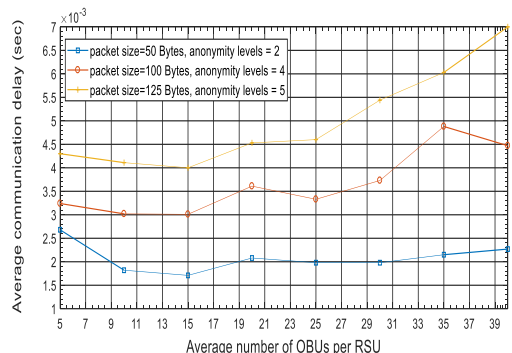

(c)

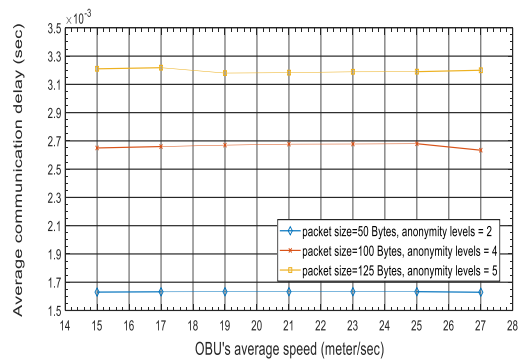

(d)

Fig. 14: Average communication delays and average packet loss ratio under various anonymity levels (α=2, α=4, and α=5)

$n=50$, randomly chosen secrets per group. We computed $P_L$ with various numbers of ZKP proofs, where $\mu = 5, 6, 8$, and 10. A higher number of ZKP proofs contributes to better resiliency against an RSU cheater, but at the same time introduces extra latency and communication overhead, which might not be suitable for time sensitive VANET's services. Therefore, we provide an authentication protocol that is customizable and capable of dynamically updating its internal state based on service preferences (reliability versus latency) and user's privacy settings.

We further, analyze the probability, $q$ of false authentication, where two different OBUs within the same OBU-group produce the same sets of $\mu$ ZKP proofs for RSU verification and authentication. As depicted in Fig. 13, the probability, $q$ was estimated with ($n = 15$, $\mu =5$) and various values for $h = 4, 5, 6,$ and 7. As the number of iterations per ZKP proof increases, $q$ decreases. Also, there is an exponential decrease in the probability of false authentication as the number of randomly selected secrets grows from 5 secrets per OBU to 15 secrets per OBU.

## X. SIMULATION RESULTS OF AGZKP-AP

To analyze the performance of AGZKP-AP in terms of communication delays and packet loss, we build a VANET simulation environment based on OMNET++. The proposed AGZK-AP was implemented on each OBU and RSU node deployed over the simulated network. In our simulation, we considered a VANET that consist of 10 RSUs and up to 50

OBUs per RSU, performing anonymous authentication simultaneously to access the network. RSUs were distributed uniformly across VANET, with 900m apart from each other. We assumed, that each OBU node is capable of sending and receiving data if it is within a 500m communication range of an RSU. We tested the performance of AGZKP-AP under various anonymity levels parameters {α=2, α=4, α=5} and pertained simulation data related to packet loss ratio and average communication delays. In our analysis, we used authentication's packet size as another performance metric. Since anonymity level is directly proportional to the size of the authentication packets, we consider different simulation scenarios: (i) authentication packet size = 50Bytes, α=2 (ii) authentication packet size = 100Bytes, α=4, and (iii) packet size=125Bytes, α=5. For each anonymity level, a set of 48 simulation data points were collected to observe the AGZKP-AP behavior in terms of average packet loss. Also, a set of 48 simulation data points related to average communication delays were collected. Figure 14a shows the simulation results of average packet loss ratios versus the average number of OBUs per RSU. As authentication packet size increases with the respect to α from 50 Bytes to 125 Bytes, there are slight increases in the average packet loss ratios. Meanwhile, as the average number of OBU per RSU increases from 5 OBUs to 40 OBUs, the figure shows a linear increase in the average packet loss ratios under various (α)s.

Figure 14b, presents the average packet loss ratio versus the OBU's average speed (m/sec), as we elevate the protocol's anonymity level, α from 2 to 5, the average packet loss ratio increases. As OBUs' average speed increases from 14 (m/sec) to 27 (m/sec), the average packet loss ratios remain within 0.00626 and 0.0197 for various α values. However, When OBU's average speed is between 25(m/sec) and 27(m/sec), the average packet loss ratio increases from 0.00765 to 0.0765 for packet size (50 Bytes, α =2). Figure 14c shows the average communication delays versus the average number of OBUs per RSU. The average communication delays increase linearly as the number of OBUs per RSU increases. In the same time, as we increase the protocol's anonymity level, average communication delays increase due to an increase in the amount of authentication data being exchanged between an OBU and an RSU. Finally, Figure 14d presents average communication delays versus OBU's average speed (m/sec). As illustrated in figure 14d, the average communication delays remain approximately steady at 0.00163, 0.00265, and 0.00321 with α=2, α=4, α=5 respectively.

## XI. CONCLUSIONS

Protecting user's privacy and minimizing traceability are important issues that have been not considered in existing research when designing authentication protocols for VANET. Traditional authentication schemes were developed in the past to provide a secure and reliable method for validating user's credentials over the network. Vehicle Identification Number (VIN) along with data location can be used to track a user over the network. Trusted entities on the VANET can utilize authentication parameters for profiling user behaviors over such networks without the user's approval. We consider such an act as a violation of user privacy. The main issue is related to the disclosure of the user's authentication parameters when accessing the network. Therefore, we proposed a novel authentication technique called Adaptive Group-based Zero Knowledge Proof-Authentication Protocol (AGZKP-AP) based on a hybrid approach, combining common verifiable scheme with ZKP protocol to minimize the disclosures of user authentication parameters while accessing the network. Our approach is adaptive when it comes to offering various user privacy settings. The proposed protocol is integrated with a security feature that enables a network to dynamically adjust the internal state of the protocol to fit a service requirement. The amount of private information enclosed during the authentication phase can be controlled by a dynamically updated parameter, α, which serves as a threshold value for validating a user on the network. Another performance metric that controls the resiliency of AGZKP against an RSU cheater and a ZKP simulator is the number of ZKP proofs required for user verification $\mu$. We evaluate our protocol with various values for $\mu$, $n$, $h$, and $k$ and determine its resiliency against two types of threat models (RSU cheater and ZKP simulator). As illustrated in the previous section AGZKP-AP provides substantial resistance to attacks with a probability of false authentication approaching zero.

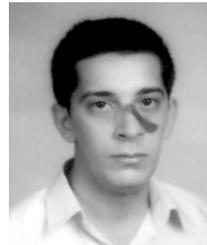

Dr. Amar Rasheed is an Assistant Professor in the Department of Computer Science at Georgia Southern University. He worked previously as a post-Doctoral Fellow in the Information Science and Technology Division of the Applied Research Laboratory (ARL) at Pennsylvania State University. His research interests include sensor modeling and data collection algorithms, efficient data collection schemes for wireless sensor networks, energy-efficient sensor data gathering mechanisms, secure mobile sensor data communication models design, cybersecurity systems, cybersecurity risk assessment and analysis, secure wireless sensor network, key pre-distribution schemes for randomly distributed sensor network, applied crypto, and the development of energy-efficient schemes for power-limited devices.

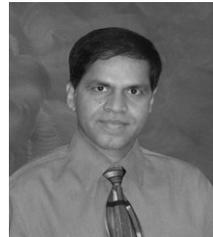

Dr. Rabi Mahapatra is a Professor in the Department of Computer Science and Engineering at Texas A&M University, USA. He was a faculty at Indian Institute of Technology, Kharagpur, India and a Faculty Fellow at IBM T.J Watson Research Center, USA. His principal areas of research are Embedded Systems, System on Chip, Low-power system design, and Data Analytic Accelerators. Dr. Mahapatra directs the Embedded System Codesign Research Group at Texas A&M University. He has been in the Editorial boards of ACM Transactions on Embedded Computing, IEEE Transactions on Parallel Distributed Systems, and EUROSIP Journal on Embedded Systems. He has published three books and more than 160 research articles in the refereed International Journals and Conference Proceedings. Dr. Rabi Mahapatra is a Ford Fellow, BOYS-CAST Fellow, Senior-Member IEEE Computer Society, and was a Distinguished Visitor of IEEE Computer Society. Dr. Mahapatra received the undergraduate Teaching Excellence award in 2010.

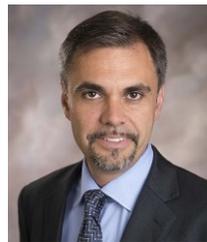

Dr. Felix G. Hamza-Lup is Professor of Computer Science at the College of Engineering and Computing, Georgia Southern University, where he is directing the Network Enabled Work-Spaces research laboratory. Dr. Hamza-Lup's research interests are in fields of Human Cognition, Human-Computer Interaction, Haptics, Web3D, and Web Security. He currently is involved with research and development on multimodal training and simulation systems for minimally invasive surgery, intelligent dialogue in e-learning systems, and autonomous sensors systems security for data aggregation. He has published more than 100 research articles in refereed International Journals and Conference Proceedings Dr. Hamza-Lup is the two time awardee of the Fulbright fellowship for outstanding research (2013, 2015) and his projects were awarded grants from several organizations including NASA SBIR, MD Anderson Cancer Foundation, NSF and the European Union research programs (Erasmus+).